\documentclass{PoS}

\title{Simple explanation of strong suppression of fermionic EFT operators \& custodial symmetry breaking}

\ShortTitle{Suppression of fermionic EFT operators \& custodial symmetry breaking}

\author{\speaker{Adolfo Guevara},$^a$ Fernando Alvarado$^a$
and Juan Jos\'e Sanz Cillero$^a$\\
\llap{$^a$}Departamento de F\'isica Te\'orica and IPARCOS\\
Universidad Complutense de Madrid, Plaza de las Ciencias 1, 28040 Madrid, Spain\\
E-mail: \email{adguevar@ucm.es}, \email{falvar03@ucm.es},
\email{jjsanzcillero@ucm.es}}

\abstract{The present approach relies on the SM chiral symmetry breaking pattern $SU(2)_L\otimes SU(2)_R\to SU(2)_{L+R}$, 
 with the EW Goldstone bosons given in a non-linear realization and the Higgs boson described by an EW 
 singlet field. In addition, we assume the presence of new physics heavy states around the TeV scale that 
 do not couple to the SM fermions, only to the SM bosonic sector. However, the mixing between gauge bosons 
 and BSM resonances induces a small indirect interaction between the BSM sector and the SM fermions. 
 This leads to an important suppression of the fermionic operators in the low-energy EW effective theory 
 (bilinear and four-fermion operators) in comparison with the purely bosonic ones. This naturally explains 
 the strong experimental bounds on fermionic operators and why these resonances could not be yet detected: 
 even if energies of the order of the TeV can be reached in present and future accelerators, their production 
 from initial SM fermions yields a very small cross section because of this suppression mechanism. On the other 
 hand, they can leave an imprint in SM boson measurements accessible to future experimental runs (e.g., the 
 oblique $S$ and $T$ parameters). Finally, we compare our results with constraints from collider data.}

\FullConference{7th Annual Conference on Large Hadron Collider Physics - LHCP2019\\
		20-25 May, 2019\\
		Puebla, Mexico}

 \usepackage{mathtools}
\usepackage{slashed}

\begin{document}

\section{Introduction}

 \footnote{This contribution is based on our work \cite{us}.}
 The search for physics beyond the Standard Model (BSM) has put very stringent limits on the masses of spin 1 particles coupling to Standard Model (SM) fermions
 \cite{Aaboud:2017yvp,Sirunyan:2017ygf,Schael:2013ita}. These bounds are obtained using operators with canonical dimension 6 such as Fermi-like operators, this is, 
 four-fermion contact interactions. The inverse mass scale parameter in the Lagrangian density is often taken to be the mass of such BSM field. 
 The problem with this approach is that such field may not couple directly to the SM fermions, which might found additional suppression due to the need of 
 a mediator between these fields. If one relies on effective Chiral Lagrangians to recover the SM operators and regard the BSM spin 1 fields as resonances that 
 do not couple directly to the SM fermions, a natural suppression to observables involving fermions comes about due to a need for additional exchange of 
 SM gauge bosons. In this work we show that these rather stringent experimental limits may not necessarily imply such restrictive bounds on the masses of the resonances, by means 
 of a somewhat naive estimation of such masses. We also obtain a more elaborate estimation of the masses of spin 1 resonances using experimental bounds on 
 the oblique $S$ and $T$ parameters \cite{Peskin:1990zt,Peskin:1991sw,Altarelli:1990zd,Altarelli:1991fk}.\\

 Considering the electroweak sector of the Standard Model (SM) to be described by an effective theory of a more fundamental strongly-coupled theory, where 
 its scalar sector symmetry group 
 can be generalized to that of the Electroweak Effective Theory\footnote{The Electroweak Effective Field Theory is also called Higgs Effective Field Theory (HEFT)
 or Electroweak Chiral Lagrangians.} (EWET),
 the Electroweak Chiral Symmetry group $G=SU(2)_L\otimes SU(2)_R$. Relying on this symmetry and without adding more fields than those of 
 the SM, an effective field theory at {\it low energies} of such strongly-coupled theory can be developed to describe the interactions of these fields. The 
 procedure is then completely analogous to that of Chiral Perturbation Theory ($\chi$PT) \cite{Weinberg:1978kz,Gasser:1983yg,Gasser:1984gg}: 
 The Electroweak Chiral Symmetry gets spontaneously broken to the custodial symmetry, 
 $H= SU(2)_{L+R}$, which generates three Goldstone bosons. Afterwards and in order to recover the SM interactions, the third generator of 
 $SU(2)_R$ is identified as the SM hypercharge field, while the other two are regarded as non-propagating external currents. This gives an explicit symmetry breaking 
 to $SU(2)_L\otimes U(1)_Y$. It is worth to notice that the physical Higgs field is a scalar under the whole Electroweak Chiral Symmetry group. Another key ingredient of the  
 theory is the non-linear realization of the EW Goldstone bosons. This leads to a theory resembling $\chi$PT and that can be developed in an analogous way and 
 expanded in chiral powers. With the chiral counting given in ref. \cite{Pich:2016lew,Krause:2018cwe}, the SM operators will be given by the lowest chiral dimension 
 Lagrangian density $\hat d=2$.\\

 Just as $\chi$PT can be naturally extended to include meson resonances \cite{Ecker:1988te,Ecker:1989yg}, 
 so can the EWET \cite{Pich:2016lew,Krause:2018cwe} be extended to include resonances. From the resonances Lagrangian the equations of motion (EoM) 
 are obtained and then substituted to obtain the {\it low-energy} contribution to the couplings of the EWET theory Lagrangian density of chiral 
 dimension $\hat d=4$. This has been done previously in \cite{Pich:2016lew}. However, the novelty in our calculation is that we take into account the 
 NLO contribution to the EoM of the resonances, obtaining new effects such as, {\it eg}, a non-vanishing contribution to the oblique $T$ parameter. Since LEP \cite{Schael:2013ita}
 gives the most precise data for the $S$ and $T$ parameters, we use this data set to obtain bounds on the polar-vector mass and in the axial- to 
 polar-vector resonance masses ratio.

 \section{Electroweak Effective Theory}
 
 The Electroweak Effective Theory is based in the spontaneous symmetry breaking of the Electroweak Chiral Symmetry $G=SU(2)_L\otimes SU(2)_R$ to the 
 custodial symmetry $H=SU(2)_{L+R}$. If the four fundamental scalar fields of the SM are collected in a $2\times2$ matrix \cite{Pich:2016lew}
 \begin{equation}
  \Sigma = \left(\begin{array}{cc}{\Phi^0}^*&\Phi^+\\\hspace*{-1.7ex}-\Phi^-&\Phi^0\end{array}\right),
 \end{equation}
 the Lagrangian of these fields is given by 
 \begin{equation}
  \mathcal{L}(\Phi)=\frac{1}{2}\left\langle\left(D^\mu\Sigma\right)^\dagger D_\mu\Sigma\right\rangle
  -\frac{\lambda}{16}\left(\left\langle\Sigma^\dagger\Sigma\right\rangle-v^2\right)^2,
 \end{equation}
 where 
 \begin{equation} 
 D_\mu\Sigma=\partial_\mu\Sigma+ig\frac{\overrightarrow\sigma}{2}\cdot\overrightarrow W_\mu\Sigma
 -ig'\Sigma\frac{\overrightarrow \sigma}{2}\cdot\overrightarrow B_\mu,
 \end{equation}
 $v=246$ GeV is the EW vacuum expectation value, $g$ and $g'$ the gauge group couplings and $\sigma_i$ the Pauli matrices.
 Instead of representing the scalar sector of the SM in cartesian components (linear representation), it is given in 
 a polar representation\footnote{This non-linear representation of the Goldstone bosons is also found in the Standard Model when 
 one {\it gauges them out} to become the longitudinal parts of the gauge bosons (unitary gauge).} 
 (non-linear representation) as follows 
 \begin{equation}
  \Sigma(x) = \frac{1}{\sqrt{2}}\left[v+h(x)\right]U(\varphi(x)),
 \end{equation}
 where $h$ is the physical Brout-Englert-Higgs (BEH) field and $U$ transforms under $G$ as
 \begin{equation}
  U(\varphi)=\exp\left\{i\overrightarrow\sigma\cdot\frac{\overrightarrow\varphi}{v}\right\}\xrightarrow{\hspace*{1ex}G\hspace*{1ex}} 
  g_LU(\varphi)g_R^\dagger\hspace*{20ex} g_{L,R}\in SU(2)_{L,R}.
 \end{equation}
 This means that $\mathcal{L}(\Phi)$ can be written as 
 \begin{equation}
  \mathcal{L}(\Phi)=\frac{v^2}{4}\left\langle\left(D^\mu U\right)^\dagger D_\mu U\right\rangle 
  + \frac{1}{2}\left(\partial_\mu h\hspace*{.5ex}\partial^\mu h\right) - V(h).
 \end{equation}
 In the unitary gauge $U=1$, and one recovers the SM values for the masses of the gauge bosons in terms of the {\it vev}, $g$ and $g'$. This 
 procedure is, however, equivalent to selecting a definite non-linear representation of the coset $G/H$ coordinates 
 $(u_L(\varphi),u_R(\varphi))\to(U(\varphi),1)$, where 
 $U(\varphi)=u_L(\varphi)u_R^\dagger(\varphi)$, and the general coordinates transform as \cite{Coleman:1969sm,Callan:1969sn} 
 \begin{equation}
  u_{\chi}(\varphi)\xrightarrow{\hspace*{1ex}G\hspace*{1ex}} g_\chi u_{\chi}(\varphi) g_h^\dagger(\varphi,g),
 \end{equation}
 for $\chi=R,L$, where $g=(g_L,g_R)\in G$.
 A more convenient representation of the Goldstone fields for the extension of the model (when the resonances are included), is obtained for $u(\varphi)=u_L(\varphi)=u_R^\dagger(\varphi)$, 
 where
 \begin{equation}
 u(\varphi)=\exp\left\{i\overrightarrow\sigma\cdot\frac{\overrightarrow\varphi}{2v}\right\}.
  \end{equation}
  Tensors transforming as triplets under the vector subgroup $H$, $R\to g_h Rg_h^\dagger$, have covariant derivatives
  \begin{equation}
   \nabla_\mu R=\partial_\mu R+ \left[\Gamma_\mu,R\right]\hspace*{3.5ex}\xrightarrow{\hspace*{1ex}G\hspace*{1ex}}\hspace*{3.5ex} g_h \nabla_\mu Rg_h^\dagger,
  \end{equation}
  where the connection is
  \begin{equation}
   \Gamma_\mu=\frac{1}{2}\left(\Gamma^L_\mu+\Gamma^R_\mu\right),\hspace*{4ex}
   \Gamma^\chi_\mu=u^\dagger_\chi(\varphi)\left(\partial_\mu-i\hat \chi_\mu\right)u_\chi(\varphi)
  \end{equation}
 being $\hat \chi_\mu=\hat W_\mu, \hat B_\mu$ for $\chi=L,R$ respectively.
  Since the physical Higgs boson transforms as a scalar under the whole symmetry of the Lagrangian density, there is no restriction as to how many of such 
  fields can be considered in the Lagrangian density. Therefore, every operator in the Lagrangian density must be multiplied by a power series in $x=h/v$. 
  Thus, the scalar sector Lagrangian density, $\mathcal{L}(\Phi)$, is expressed as
 \begin{equation}
  \mathcal{L}(\Phi)=\frac{v^2}{4}\mathcal{F}_u\left\langle u_\mu u^\mu\right\rangle + \frac{1}{2}\left(\partial_\mu h\hspace*{.5ex}\partial^\mu h\right) - V,
 \end{equation}
 where 
 \begin{equation}
  u_\mu=u_\mu^\dagger=i\left(\Gamma^R_\mu-\Gamma^L_\mu\right)=i\left[u(\partial_\mu-i\hat B_\mu)u^\dagger - u^\dagger(\partial_\mu-i\hat W_\mu)u\right],
 \end{equation}
 being $\mathcal{F}_u$ and $V$ power series of $x$ whose lowest order term are $1$ and the Higgs mass term respectively.
 The part concerning the gauge bosons and their interactions are given by the familiar Yang-Mills Lagrangian. 
     \begin{equation}\label{LO Lagrangian}
     \mathcal{L}_\textrm{\scriptsize YM}= -\frac{1}{2g^2}\hat W_{\mu\nu}\hat W^{\mu\nu} - \frac{1}{2g'^2}\hat B_{\mu\nu} \hat B^{\mu\nu}, 
     \end{equation}
     where the field strength tensors are given by 
     \begin{eqnarray}
     \hat W_{\mu\nu}&=& \partial_\mu\hat W_\nu - \partial_\nu\hat W_\mu -i\left[\hat W_\mu, \hat W_\nu\right], \nonumber\\ 
     \hat B_{\mu\nu}&=& \partial_\mu\hat B_\nu - \partial_\nu\hat B_\mu -i\left[\hat B_\mu, \hat B_\nu\right].     
     \end{eqnarray}
     Here, $\hat W_\mu = -g W^i_\mu \frac{\sigma^i}{2}$ where $\sigma^i$ are the $SU(2)_L$ Pauli matrices and similarly $\hat B_\mu = -g' B^i_\mu \frac{\sigma^i}{2}$ 
     for the $SU(2)_R$ group. The fermionic sector of the EWET is given by the Lagrangian density
         \begin{equation}\label{Fermions}
     \mathcal{L}_\psi= i\overline{\psi}\slashed{D}\psi -v^2\langle J_S\rangle,
   \end{equation}
  where $J_S=J_Y + J_Y^\dagger$, being $J_Y=(1/v)\overline{\xi}_L \mathcal{Y}\xi_R$ and $\mathcal{Y}$ a power series in $x$ whose order zero term 
  gives the SM Yukawa couplings. Here, the covariant derivative is given by the following expression
  \begin{equation}\label{Cov Der psi}
   D_\mu^{\chi} \psi_{\chi}= \left(\partial_\mu - i\hat \chi_\mu - i \hat X_\mu\frac{\mathfrak{B}-\mathfrak{L}}{2}\right)\psi_\chi, 
  \end{equation}
  where $\hat\chi_\mu=\hat B_\mu,\hat W_\mu$ for $\chi=R,L$, being $\psi_{R/L}=\frac{1}{2}(1\pm\gamma_5)\psi$ the chirality projection of the Dirac spinor and the $\mathfrak{B}$ and $\mathfrak{L}$ 
  operators that act on the fermion fields, whose eigenvalues are the barion and lepton numbers respectively. The chiral projections of the fermion fields 
  transform under $G$ as 
  \begin{equation}
   \psi_\chi\to g_\chi \psi_\chi.
  \end{equation}
  A spinor transforming under the custodial symmetry group $SU(2)_{L+R}$ can be obtained multiplying the left (right) Weyl spinor by $u_{L/R}^\dagger$
  \begin{equation}
   \xi_\chi = u_\chi^\dagger \psi_\chi,\hspace*{6ex} \overline{\xi}_\chi = \overline{\psi}_\chi u_\chi,
  \end{equation}
  so that one gets a new spinor which transform under $G$ as $\xi_\chi\to g_h\xi_\chi$, with $g_h\in SU(2)_{L+R}$. 
  To recover the Standard Model, the coupling between the $\hat X$ field and the SM fermions 
  given in eq. (\ref{Cov Der psi}) must be such that $\hat X$ is the SM hypercharge field,
   \begin{equation}
    \hat X_\mu = -g'B_\mu.
   \end{equation}
  For future reference, we define the left and right fermion currents as
  \begin{subequations}
   \begin{align}
   \left(J^\ell_\mu\right)_{mn} &= \frac{1}{2}{\overline{\xi}_n\gamma_\mu(1-\gamma_5)\xi_m},\\
   \left(J^r_\mu\right)_{mn} &= \frac{1}{2}{\overline{\xi}_n\gamma_\mu(1+\gamma_5)\xi_m},
   \end{align}
  \end{subequations}
  where $m$ and $n$ are flavor indices transforming under $SU(2)_{L+R}$.

\section{Equations of motion for SM bosons}\label{section:EoM}
   The equations of motion for the Goldstone bosons are given by the following expressions
   \begin{equation}\label{EoM GB}
    \nabla_\mu u^\mu = -u^\mu \partial_\mu\left(\frac{h}{v}\right)\frac{\mathcal{F}_u'}{\mathcal{F}_u}-\frac{2}{\mathcal{F}_u}J_P+\frac{1}{\mathcal{F}_u}\langle J_P\rangle,
   \end{equation}
 where $J_P = i(J_Y - J_Y^\dagger)$ and the prime on the functions of the BEH boson means that it is derived with respect to $x=h/v$. For the BEH field we have
 \begin{equation}\label{EoM Higgs}
  \Box \left(h/v\right) = \frac{1}{4}\mathcal{F}_u'\langle u_\mu u^\mu\rangle - V' -\langle J_S'\rangle.
 \end{equation}
 
 For the gauge boson fields we find  
 \begin{subequations}
  \begin{align}
  D^\nu \hat W_{\nu\mu,i} &= \frac{(gv)^2}{4}\mathcal{F}_u\langle u_\mu \sigma_i^L\rangle 
  + \frac{g^2}{2}\langle \xi^{\alpha}_L\overline{\xi}^\beta_L\gamma_\mu^{\beta\alpha} \sigma_i^L\rangle,\label{EoM W}\\
  D^\nu \hat B_{\nu\mu,i} &= -\frac{(g'v)^2}{4}\mathcal{F}_u\langle u_\mu \sigma_i^R\rangle 
  + \frac{g'^2}{2}\langle \xi^{\alpha}_R\overline{\xi}^\beta_R\gamma_\mu^{\beta\alpha} \sigma_i^R\rangle - 
  (g')^2 \overline{\psi}\gamma_\mu\frac{\mathfrak{B}-\mathfrak{L}}{2}\delta_{3,i}\psi,\label{EoM Bs}
  \end{align}\label{EoM gauge}
 \end{subequations}
 where $\beta$ and $\alpha$ are spinor indices, $\delta_{3,i}$ is the Kronecker delta, $\sigma_i^\chi=u_\chi^\dagger \sigma_i u_\chi$ for $\chi = R,L$
 and $D^\nu \hat X_{\nu\mu,i}=\langle D^\nu \hat X_{\nu\mu}\sigma_i\rangle$. 
 Nevertheless, it must be noticed that the only physical equation of motion in eqs. (\ref{EoM Bs}) is that of $\hat B^3$. 
 When the $B^1$ and $B^2$ fields vanish, this gives
 \begin{equation}\label{EoM B-Hyp}
  \partial^\nu  B_{\nu\mu}=-\frac{(g'v)^2}{4}\mathcal{F}_u\left( u_L u_\mu u_L^\dagger\right)_3 
  + \frac{g'^2}{2}\left( \psi^{\alpha}_L\overline{\psi}^\beta_L\gamma_\mu^{\beta\alpha}\right)_3 - 
  \frac{(g'v)^2}{2} {J_X}_\mu,
 \end{equation}
 where $B_{\mu\nu}=\partial_\mu \hat B^3_\nu - \partial_\nu \hat B^3_\mu$ and $J_X^\mu=\overline{\xi}_m\gamma^\mu\frac{\mathfrak{B}-\mathfrak{L}}{v^2}\xi_m$,
 being $m$ the $SU(2)_{L+R}$ index of the spinors and a sum over repeated indices is implied, meaning that the current $J_X$ 
 behaves as a scalar under the chiral group. It must be noticed that any contribution of the fields $\hat B^1$ and $\hat B^2$ must be neglected 
 in the calculation of vertex functions since they are regarded as external currents, instead of fundamental fields and, therefore, are to be considered 
 to vanish. However, these fields are needed in order to use relations allowing us to express eqs. (\ref{EoM gauge}) as operators transforming under 
 the residual custodial symmetry. Thus, the covariant derivatives in eq.(\ref{EoM gauge}) can be expressed as
  \begin{subequations}
  \begin{align}
  D^\nu \hat W_{\nu\mu,i} &=\langle\nabla^\nu f^L_{\nu\mu}\sigma_i^L\rangle -
  \frac{i}{2}\langle \left[f^L_{\nu\mu},u^\nu\right]\sigma_i^L\rangle,\label{Dnu W}\\
  D^\nu \hat B_{\nu\mu,i} &=\langle\nabla^\nu f^R_{\nu\mu}\sigma_i^R\rangle + \frac{i}{2}\langle\left[f^R_{\nu\mu},u^\nu\right]\sigma_i^R\rangle,\label{Dnu B}
    \end{align}\label{Cov Deriv}
  \end{subequations}
  where $f^{L/R}_{\mu\nu}=\frac{1}{2}({f_+}_{\mu\nu}\pm{f_-}_{\mu\nu})$, where
  \begin{equation}
   f^{\mu\nu}_\pm=u^\dagger\hat W^{\mu\nu} u \pm u\hspace*{0.5ex} \hat B^{\mu\nu} u^\dagger.
  \end{equation}
 Putting together eqs. (\ref{EoM W}), (\ref{EoM Bs}) for $i=3$ and (\ref{Cov Deriv}) we find 
  the following relations
  \begin{subequations}
   \begin{align}
    \nabla^\nu f^L_{\mu\nu} = & \frac{i}{2} \left[f^L_{\mu\nu},u^\nu\right]
    - \frac{(gv)^2}{4}\mathcal{F}_u u_\mu - \frac{g^2}{2}\xi^{\alpha}_L\overline{\xi}^\beta_L\gamma_\mu^{\beta\alpha}
    +\frac{g^2}{4}\langle\xi^{\alpha}_L\overline{\xi}^\beta_L\gamma_\mu^{\beta\alpha}\rangle,\label{eq Nabla fL}\\
    \nabla^\nu f^R_{\mu\nu} = & \left\langle\left(- \frac{i}{2} \left[f^R_{\mu\nu},u^\nu\right]
    + \frac{(g'v)^2}{4}\mathcal{F}_u u_\mu - \frac{\left(g'\right)^2}{2}\xi^{\alpha}_R\overline{\xi}^\beta_R\gamma_\mu^{\beta\alpha}\right)\sigma_3^R\right\rangle\frac{\sigma_3^R}{2} \nonumber\\
    & + \frac{(g'v)^2}{2}{J_X}_\mu \frac{\sigma_3^R}{2}+ \widetilde{\nabla^\nu f^R_{\mu\nu}},\label{eq Nabla fR}
   \end{align}\label{EoM f}
  \end{subequations}
  where $\widetilde{\nabla^\nu f^R_{\mu\nu}}=\sum_{i=1,2}\left\langle{\nabla^\nu f^R_{\mu\nu}}\sigma_i^R\right\rangle\frac{\sigma_i^R}{2}$.
  One of the main advantages of formulating the SM using the chiral symmetry is that all the SM is given by the most general Lagrangian 
  with chiral dimension $\hat d=2$, meaning that NLO operators will encode only BSM interactions. 

\section{Resonance contribution to low energy constants}\label{section:Resonance EoM}

 \subsection{Equations of motion of spin 1 resonances}
  In an analogous way to the extension of $\chi$PT which incorporates the meson resonances as active degrees of freedom \cite{Ecker:1988te,Ecker:1989yg}, one can 
 include resonances to obtain their contribution to the couplings of the $\mathcal{O}(p^4)$ chiral Lagrangian operators \cite{Pich:2016lew,Krause:2018cwe}. Thus, one 
 must obtain the equations of motion of the resonances and substitute the expressions for the classical fields in the Lagrangian for these resonances. 
 In the antisymetric formalism \cite{Ecker:1989yg}, such Lagrangian for spin = 1 resonances is
 \begin{equation}\label{eq:Res_Lagrangian}
  \mathcal{L}_{R} = -\frac{1}{2}\left\langle \nabla^\lambda R_{\lambda\mu}\nabla_\sigma  R^{\sigma\mu}-\frac{1}{2}M_R^2 R_{\mu\nu}  R^{\mu\nu}\right\rangle + 
  \left\langle  R_{\mu\nu}\chi^{\mu\nu}_{R} \right\rangle \hspace*{5ex} \left( R =  V,  A\right),
 \end{equation}
 The definitions of $\chi_{ R}^{\mu\nu}$ for $R = V, A$ are given as follows \cite{Pich:2016lew,Krause:2018cwe}
 \begin{subequations}
  \begin{align}
   \chi_{ V}^{\mu\nu} & = \frac{F_{ V}}{2\sqrt{2}}f_+^{\mu\nu} + \frac{\widetilde F_{ V}}{2\sqrt{2}}f_-^{\mu\nu} + 
   i\frac{G_{ V}}{2\sqrt{2}}\left[u^\mu,u^\nu\right] + \frac{\widetilde\lambda_1^{h V}}{\sqrt{2}}\left[\left(\partial^\mu h\right)u^\nu - 
   \left(\partial^\nu h\right)u^\mu\right],\\
   \chi_{ A}^{\mu\nu} & = \frac{\widetilde F_{ A}}{2\sqrt{2}}f_+^{\mu\nu} + \frac{ F_{ A}}{2\sqrt{2}}f_-^{\mu\nu} + 
   i\frac{\widetilde G_{ A}}{2\sqrt{2}}\left[u^\mu,u^\nu\right] + \frac{\lambda_1^{h A}}{\sqrt{2}}\left[\left(\partial^\mu h\right)u^\nu - 
   \left(\partial^\nu h\right)u^\mu\right],
  \end{align}\label{Resonance couplings}
 \end{subequations}
 where the couplings with tilde mean that the corresponding operator in the Lagrangian \ref{eq:Res_Lagrangian} do not conserve parity.
 Thus, the axial-vector resonance contribution to the EWET Lagrangian couplings can be obtained from the polar-vector ones by making the substitutions 
 $F_{ V}\to \widetilde F_{ A}$, $\widetilde F_{ V}\to F_{ A}$, $\widetilde \lambda^{h V}_1\to \lambda^{h A}_1$ and
 $G_{ V} \to \widetilde G_{ A}$.  Notice that all coupling constants have canonic dimension 1 except  
 for $\widetilde \lambda_1^{h V}$ and $\lambda_1^{h A}$, which are canonically dimensionless. The equations of motion for the resonances are given 
 by the following expression \cite{Ecker:1988te}
 \begin{equation}
  \nabla^\mu\nabla_\sigma R^{\sigma\nu}-\nabla^\nu\nabla_\sigma R^{\sigma\mu}+M_R^2R^{\mu\nu}=-2\chi^{\mu\nu}_R
 \end{equation}
 with which the classical field expressions for the resonance fields are given by
 \begin{equation}
  R^{\mu\nu}_{class}=-\frac{2}{M_R^2}\chi^{\mu\nu}+\frac{2}{M_R^2}\left[\nabla^\nu\nabla_\sigma\left(\frac{\chi^{\mu\sigma}_R}{M_R^2}\right)
  -\nabla^\mu\nabla_\sigma\left(\frac{\chi^{\nu\sigma}_R}{M_R^2}\right)\right]+\mathcal{O}\left(p^6\right)
 \end{equation}
 Substituting the previous expression in the Lagrangian for the resonance field gives the contribution to the EWET couplings of the Lagrangian
 \begin{equation}
  \mathcal{L}_R(R_{class}) = -\frac{1}{M_{\hat R}^2}\langle\chi^{\mu\nu}_R\chi_{\mu\nu,R}\rangle-2\left\langle\nabla_\nu\left(\frac{\chi^{\mu\nu}_R}{M_R^2}
  \right)\nabla^\sigma\left(\frac{\chi_{\mu\sigma,R}}{M_R^2}\right)\right\rangle+\mathcal{O}(p^8),
 \end{equation}
 where the term with the covariant derivatives is the novel contribution to the EoM, which considering the gauge and Goldstone fields EoM
 gives, among other effects, a custodial symmetry breaking term contributing to the oblique $T$ parameter \cite{us,Pich:2016lew,Herrero:1993nc}.
 
\section{Phenomenology}

 Using the equations of motion for the Goldstone and gauge bosons in $\mathcal{L}_R(R_{class})$, the contribution to the couplings of the 
 EWET Lagrangian is obtained by making a comparison with the full $\mathcal{O}(p^4)$ basis given in reference \cite{Pich:2016lew,Krause:2018cwe}. Thus, the 
 four-fermion contact interaction given in the phenomenological Lagrangian in references \cite{Aaboud:2017yvp,Sirunyan:2017ygf,Schael:2013ita} 
 \begin{equation}\label{eq:phenom}
  \mathcal{L}_{EWET}\supset\mathcal{L}_{qq}=\frac{2\pi}{\Lambda^2}\left(\eta_{\ell\ell}J^\ell_\mu J^{\ell,\mu}+
 \eta_{rr}J^r_\mu J^{r,\mu}+2\eta_{r\ell}J^r_\mu J^{\ell,\mu}\right),
 \end{equation}
 can be expressed as a linear combination of the terms in $\mathcal{L}_{EWET}$. The most stringent experimental bound on the BSM scale is found for
 $\eta_{\ell\ell}=\eta_{rr}=\eta_{r\ell}=-1$, which is $\Lambda\gtrsim20$ TeV. This parameter $\Lambda$ is commonly taken to be the compositeness scale or 
 the mass of the resonance, however since in our model the resonances do not couple directly to the SM fermions, the former statement is not true.
 Comparing this with the contribution from the spin 1 resonances and 
 after using the Weinberg sum rules \cite{Weinberg:1967kj} \footnote{Here we assume parity invariance, so all the resonance couplings with tilde are set to zero. 
 Thus, we use the Weinberg sum-rules $F_V^2-F_A^2-v^2=0$ and $F_V^2 M_V^2 -F_A^2 M_A^2=0$.}
 we get 
 \begin{equation}\label{eq:4Fermion_supp}
  \frac{2\pi}{\Lambda^2}=\frac{4m_Z^4-8m_Z^2m_W^2+7m_W^4}{24v^2M_V^4}\,\,\frac{r^3+1}{r^2(r-1)},
 \end{equation}
 where $r=M_A^2/M_V^2$. Some bounds on the vector resonance mass and the ratio $r$ are given in Table \ref{tab:4fermion}, where we take $\Lambda\geq$20 TeV.
 These are, however, not to be regarded as precise predictions of the model, since this naive approach gives an estimation that does not consider correlation 
 among the $\eta_{ij}$ parameters in eq. (\ref{eq:phenom}).\\ 
 \begin{table}[!ht]
 \centering
 \begin{tabular}{|l | l |}\hline
 \hspace*{0ex}$r=M_A^2/M_V^2$ & lower bound for $M_V$ \\\hline
 \hspace*{1.5ex}$1+10^{-3}$  & \hspace*{5ex}$1.9$ TeV\\
 \hspace*{3.5ex}$1.1$  & \hspace*{5ex}$0.6$ TeV\\
 \hspace*{4.2ex}$2$  & \hspace*{5ex}$0.3$ TeV\\
 \hspace*{3.9ex}$\infty$  & \hspace*{5ex}$0.3$ TeV\\\hline
 \end{tabular}\caption{Bounds for $M_V$ as a function of the ratio $r$ using the experimental bound $\Lambda\gtrsim20$ TeV.}\label{tab:4fermion}
 \end{table}

\begin{figure}[!ht]
  \centering\includegraphics[scale=0.54]{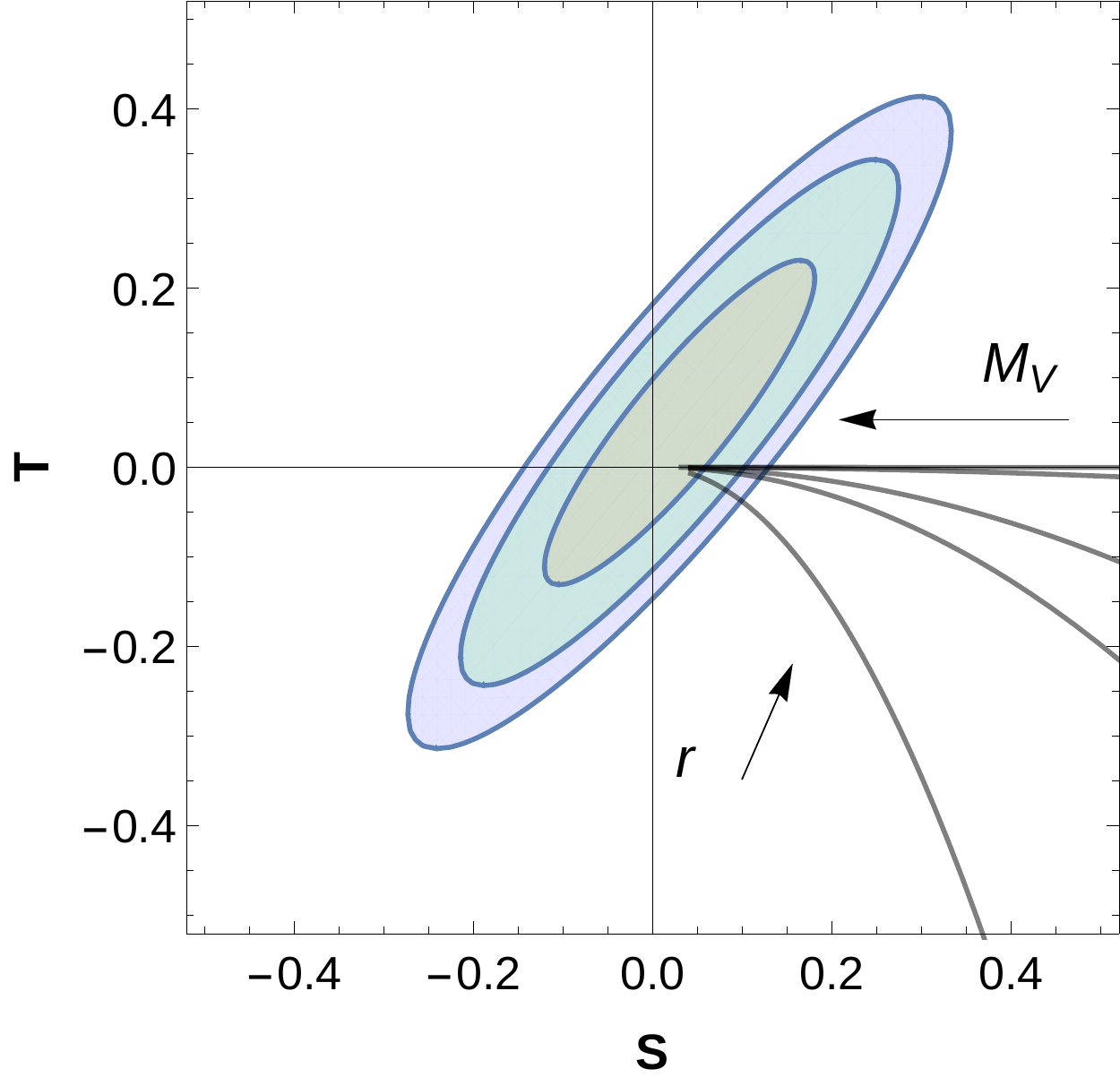}\hspace*{3ex}\includegraphics[scale=0.60]{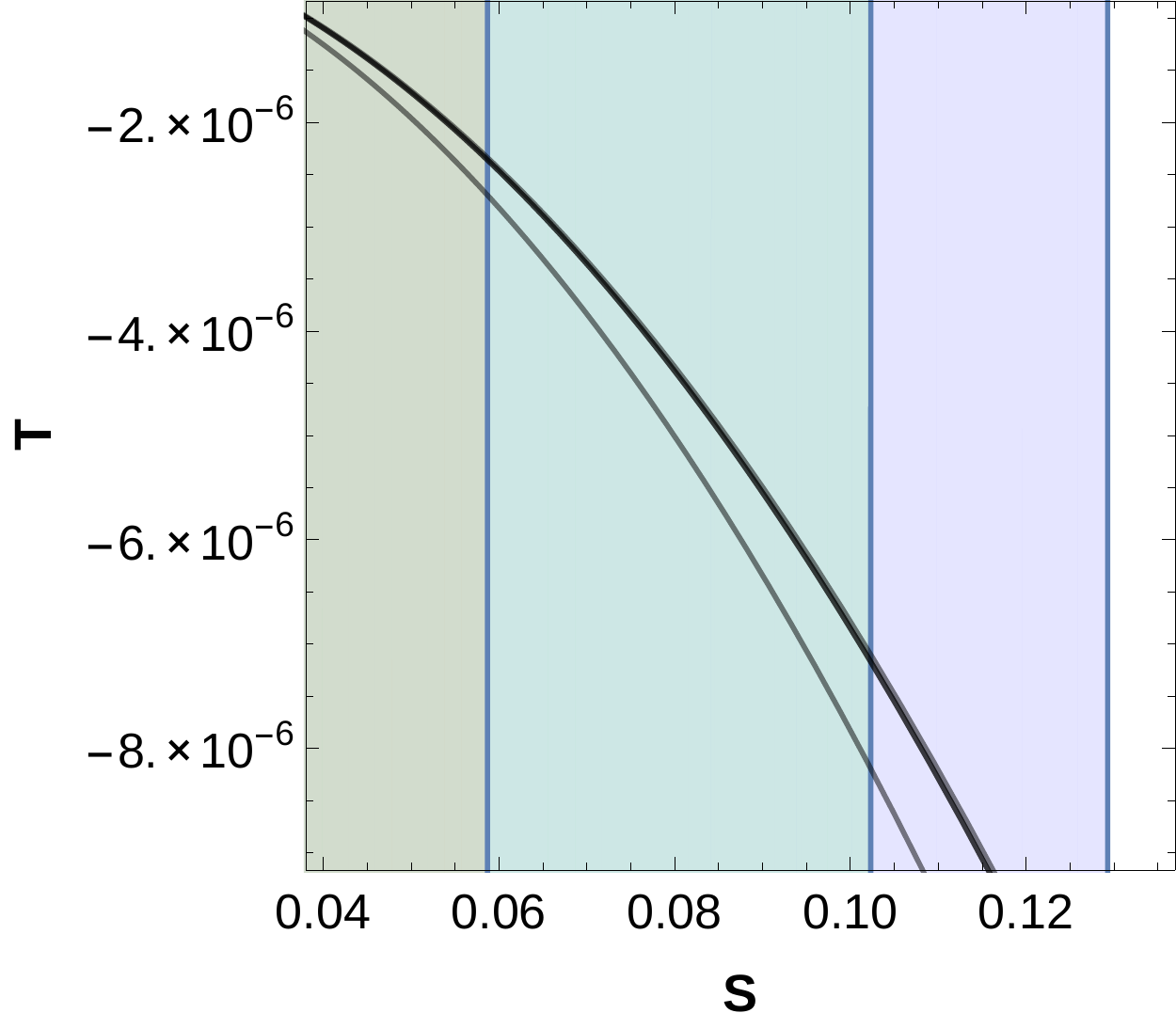}\caption{Allowed values for $T$ and $S$ 
  for LEP data at 68\% (gray), 95\% (pale green) and 99\% (violet).
  Plots for $S$ vs $T$ are given for $r-1=10^{-4},10^{-3},5\cdot10{-3}, 10^{-2}$ and $0.5$ on the left-hand-side figure. Plots for $S$ vs $T$ with a {\it zoom} 
  into the figure of the left-hand-side for $r=2,3,4,5$ on the right-hand-side. Notice that the plots converge rapidly for $r\geq3$.}\label{fig:S-T}
 \end{figure}

 To give a more elaborate estimation of the vector resonance mass as a function of the ratio 
 of spin 1 resonance masses we make use of the contributions to the $S$ and $T$ oblique parameters. The $S$ and $T$ parameters can be given as functions 
 of the couplings of the EWET Lagrangian density \cite{Herrero:1993nc}. The leading order contribution to the $S$ parameter comes from the 
 $\mathcal{O}(p^2)$ part of the EoM of the resonances \cite{Pich:2016lew,Pich:2013fea,Pich:2012dv}, at that order there is no contribution to the $T$ parameter, since 
 custodial symmetry breaking effects do not come about until $\mathcal{O}(p^4)$ terms are considered in the EoM of the spin 1 
 resonance fields \cite{us}. The expression for these parameters in terms of $v$, the gauge bosons masses and the 
 spin 1 masses are
 \begin{subequations}\label{eq:Oblique}
  \begin{align}
   S&=\frac{4\pi v^2}{M_V^2}\, \, \, \frac{r+1}{r}, \\
   T&=-\pi\frac{ v^2\left(m_Z^2-m_W^2\right)}{M_V^4}\, \, \frac{m_Z^2}{m_W^2}\, \,  \frac{r^3+1}{r^2(r-1)}.
  \end{align}
 \end{subequations}
 In the previous expressions, both Weinberg sum rules have been assumed.
 Using the limits given by LEP \cite{Schael:2013ita}, a contour plot can be drawn in the $S$-$T$ plane for the 68\%, 95\% and 99\% confidence level. 
 Thus, a limit on the vector resonance mass as a function of $r$ can be obtained for a definite confidence level. 
 The limits for the aforementioned confidence levels are depicted in Figure \ref{fig:S-T}, where plots of $S$ vs $T$ are also shown for different 
 values of $r$. 

 Bounds on the vector resonance mass can be obtained using the experimentally allowed regions for $S$ and $T$ for different confidence levels. 
 Notice that as $r$ grows, $T$ converges rapidly to the limit 
 \begin{equation}
  \lim_{r\,\to\,\infty} T = -\pi\frac{ v^2\left(m_Z^2-m_W^2\right)}{M_V^4}\, \, \frac{m_Z^2}{m_W^2}=-4.5\cdot10^{-4}\left(\frac{1\text{ TeV}^4}{M_V^4}\right).
 \end{equation}
 This effect can be noted in the right-hand-side plot of Figure \ref{fig:S-T}, where $r=3$ is sufficiently close to $r\to\infty$. The experimental limits
 given for $S$ and $T$ allow us to give bounds on $M_V$ for different values of $r$, shown in Table \ref{tab:S-T_bounds}. Notice that, although one loop 
 corrections are neglected in these estimates, they might become relevant for the T parameter in more general analyses \cite{Pich:2013fea,Pich:2012dv}.
 
 \begin{table}[!ht]
 \centering
  \begin{tabular}{|l| l l l |}\hline
 \hspace*{0ex}$r=M_A^2/M_V^2$ &\hspace*{3ex}& lower bound&for $M_V$ \\
 &&68\%CL&95\%CL\\\hline
 \hspace*{1.5ex}$1+10^{-3}$  &\hspace*{3ex}& $5.2$ TeV &$4.0$ TeV \\
 \hspace*{3.5ex}$1.1$  &\hspace*{3ex}& $5.1$ TeV &$3.9$ TeV \\
 \hspace*{4.2ex}$2$  &\hspace*{3ex}& $4.5$ TeV &$3.4$ TeV \\
 \hspace*{3.9ex}$\infty$  &\hspace*{3ex}& $3.7$ TeV &$2.8$ TeV \\\hline
 \end{tabular}\caption{Bounds on $M_V$ for different values of $r$ using the allowed region for $S$ and $T$ using the 68\% and 95\% confidence level data 
 from LEP \cite{Schael:2013ita}.}\label{tab:S-T_bounds}
 \end{table}

\section{Conclusions}

 In the framework of Chiral Electroweak Lagrangians, a simple explanation to the strong suppression of fermionic BSM operators is proposed. This 
 suppression stems from the fact that BSM operators appear due to the exchange of resonances which do not couple directly to the SM fermions. 
 This means an extra suppression due to additional exchange of gauge bosons. Thus, in the case of the four-fermion (Fermi-like) operators, the contact interaction 
 in eq. (\ref{eq:4Fermion_supp}), in addition to the naive expected value $1/M_V^2$,  carries an additional suppression factor of the form
 \begin{equation}
  \frac{4m_Z^4-8m_Z^2m_W^2+7m_W^4}{24v^2M_V^2}=9.6\cdot10^{-5}\left(\frac{\text{1 TeV}^2}{M_V^2}\right).
 \end{equation}
 As seen from Table \ref{tab:4fermion}, a stringent bound on the four-fermion parameter $\Lambda$ does not automatically imply an equally stringent bound on BSM heavy 
 states masses. We note, nevertheless, that the results in Table \ref{tab:4fermion} are just rough estimates. A full four-fermion operator analysis should be 
 performed in both the theoretical and experimental sides.\\
 
 In order to give more reliable bounds on the masses of BSM spin 1 fields we used the experimental limits on the oblique $S$ and $T$ parameters from LEP \cite{Schael:2013ita}.
 Considering NLO terms in the equations of motion of the resonances we obtained terms that break custodial symmetry, inducing thus a non-vanishing contribution 
 to the oblique $T$ parameter. It can be seen from eq. (\ref{eq:4Fermion_supp}) that $T$ (NLO) is suppressed with respect to $S$ (LO), as one would expect from 
 contributions actually arising at different orders in perturbation theory.\\
 
 The bounds on $M_V$ for different values of $r=M_A^2/M_V^2$ obtained in this way show that 
 BSM resonances with masses $M_R\sim1$ TeV are still compatible with experimental limits, and are also compatible with the bounds obtained through 
 fermionic observables regarding these resonances do no couple directly to the SM fermions.
 
\acknowledgments

 We thank Javier Virto for useful comments. This work was supported by the Spanish MINECO Project FPA2016-75654-C2-1-P 
and by CONACYT Project No. 250628 (`Ciencia B\'asica') and the support `Estancia Posdoctoral en el Extranjero'.

\end{document}